\documentstyle[12pt, epsfig]{article}
\newcommand{\bl} {{\bf l}}

\newcommand{\br} {{\bf r}}

\newcommand{\etahat}{{\mbox{\boldmath $\hat{\eta}$}}}

\newcommand{\meff}{M_{\mbox{\small eff}}}

\newcommand{\btab}{\begin{tabbing}}
\newcommand{\etab}{\end{tabbing}}

\newcommand{\beqn}{\begin{equation}}
\newcommand{\eeqn}{\end{equation}}
\newcommand{\barr}[1]{\begin{array}{#1}}
\newcommand{\earr}{\end{array}}
\newcommand{\beqna}{\begin{eqnarray}}
\newcommand{\eeqna}{\end{eqnarray}}
\newcommand{\btablec}{\begin{table} \begin{center}}
\newcommand{\etablec}{\end{center} \end{table}}

\newcommand{\gapproxeq}{\lower.7ex\hbox{$\;\stackrel{\textstyle>}
{\sim}\;$}}
\newcommand{\plabel}[1]{\label{#1}}
\newcommand{\pbibitem}[1]{\bibitem{#1}}
\input epsf

\begin{document}
\title{
\vspace{0.6cm}  
\Large\bf (Hybrid) Baryons: Symmetries and Masses}
\vskip 0.2 in
\author{Philip R. Page\thanks{\small \em E-mail:
prp@lanl.gov. Fax: +1-505-6671931. Tel: +1-505-6670673. Work done in collaboration
with Simon Capstick. 
} \\
{\small \em Theoretical Division, MS-B283, Los Alamos
National Laboratory,}\\ 
{\small \em Los Alamos, NM 87545, USA}}
\date{}
\maketitle
\begin{abstract}{We construct (hybrid) baryons in the flux--tube model
of Isgur and Paton. In the limit of adiabatic quark motion, 
we build proper eigenstates of orbital angular momentum and
construct the flavour, spin and $J^{P}$ of hybrid baryons
from the symmetries of the system. The lowest mass hybrid baryon is 
estimated at $\sim 2$ GeV.
}
\end{abstract}
\bigskip


PACS number(s):  \hspace{.2cm}12.38.Lg  \hspace{.2cm}12.39.Mk  \hspace{.2cm}
12.39.Pn\hspace{.2cm}12.40.Yx\hspace{.2cm}14.20.-c\hspace{.2cm} 

\section{Introduction}

Hybrid baryons are bound states of three quarks with an explicit
excitation in the gluon field of QCD. 
The construction of (hybrid) baryons in a
model motivated from the strong coupling expansion of the hamiltonian
formulation of lattice QCD, the non--relativistic
flux--tube model of Isgur and Paton \cite{paton85}, was detailed in ref. \cite{hadron}.
This model predicts the adiabatic
potentials of (hybrid) mesons at large interquark separations,
as well as the mass of the $J^{PC}=1^{-+}$ hybrid meson, 
consistent with recent estimates from
lattice QCD \cite{paton85,morning}. 
In ref. \cite{hadron} we studied the detailed flux dynamics and built
the flux hamiltonian. We restrict our discussion to cases where 
the flux settles down in a Mercedez Benz configuration (as motivated by
lattice QCD \cite{bali}).
A minimal amount of 
quark motion is allowed in response to flux motion, 
in order to work in the centre of mass frame.
Otherwise, we make the so--called ``adiabatic'' approximation, where the
flux motion adjusts itself instantaneously to the motion of the quarks.
The main result is that the lowest
flux excitation can to a high degree of accuracy (about 5\%) be simulated
by neglecting all flux--tube motions except the vibration of a 
junction. This result was obtained within the small oscillation approximation. 
The junction acquires an effective mass $\meff$ from the motion of the 
remainder of the flux--tube and the quarks.
The model is then simple: a junction is 
connected via a linear potential to the three quarks. 
The ground state of the junction motion corresponds to a conventional
baryon and the various excited states to hybrid baryons.
The junction can move in three directions, and correspondingly be excited
in three ways, giving the hybrid baryons $H_1, H_2$ and $ H_3$. 
The junction motion is depicted in Fig. \ref{rough}.

\begin{figure}[t]
\vspace{0cm}
\begin{centering}
\epsfig{file=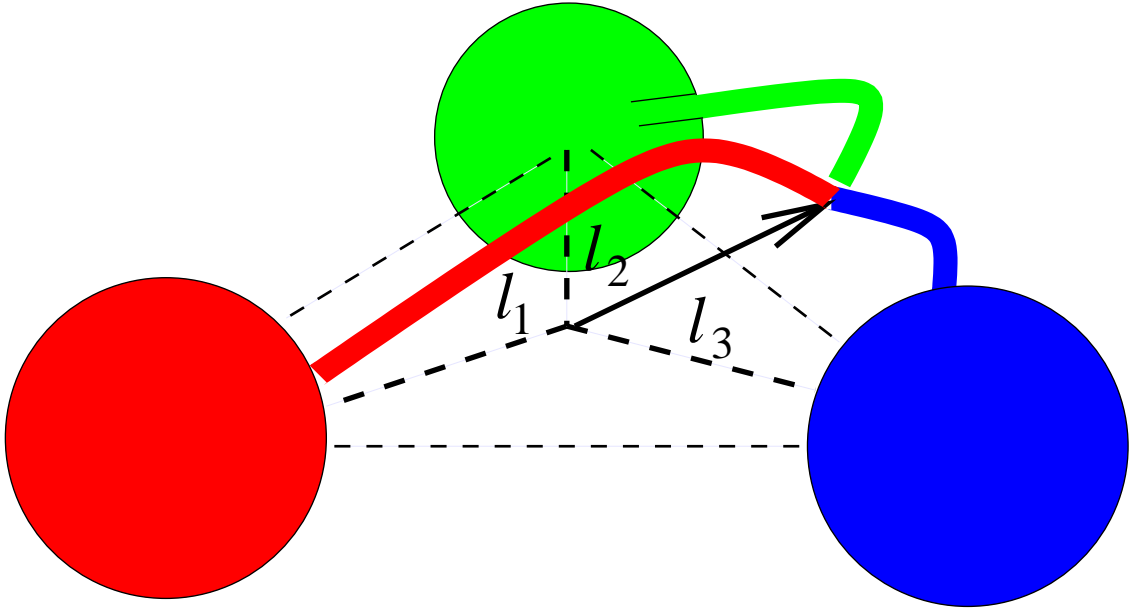,width=15cm,angle=0}
\vspace{-0.2cm}
\caption[x]{The junction connects strings coming from the three quarks. The vectors
$\br$ and $\bl_i$ respectively point from the equilibrium position 
of the junction to its current position and the quark positions.}
\plabel{rough}
\end{centering}
\end{figure}
\vspace{0.3cm}

The hamiltonian for the junction motion in the Mercedez Benz configuration
is simply the kinetic energy of the junction added to the sum of the lengths
from the junction to the quarks multiplied by the string tension $b$,

\beqna \plabel{ham3}
{H}_{\mbox{\small flux}}=\frac{1}{2}\meff\; {\bf \dot{r}^2}+
b\sum_{i=1}^{3}|\bl_i-\br|
\eeqna

We shall be taking ansatz wave functions of the form

\beqn \plabel{psih}
\etahat_{-}\cdot\br \;\;\Psi_B(\br)\hspace{1cm}
\eeqn
for $H_1$ hybrid baryons, where $\Psi_B(\br)$ is an exponential function.
It is not difficult to show that $\etahat_{-}$ lies in the plane spanned
by the three quarks (the ``QQQ plane'').

\section{Quantum numbers of low--lying hybrid baryons}

{\bf Angular Momentum:} 

The hamiltonian in Eq. \ref{ham3} is not invariant under rotations in the
junction position $\br$, with fixed quark positions. When the 
junction wave function, which is hence not an eigenfunction
 of angular momentum,
is combined with the quark motion wave functions,
which are eigenfunctions of angular momentum, it must be done in 
such a way that the total angular momentum of the junction and quark motion is
well--defined. Obtaining a well--defined total angular momentum is
a technically challenging problem that is an artifact of
 the adiabatic approximation,
which separates junction and quark motion. 
We here merely give an intuitive argument why the 
total angular momentum $L$ of the $H_1$ baryons is expected to be 1. 

The hybrid baryon wave function is proportional to $\etahat_-\cdot\br$,
and since $\etahat_-$ lies in the QQQ plane, 
it can be regarded as the x--axis, so that
$\etahat_-\cdot\br = \sqrt{\frac{2\pi}{3}} r
(-Y_{11}({\bf \hat{r}}) + Y_{1-1}({\bf \hat{r}}))$
in terms of spherical harmonics. If the mathematics of conservation of
angular momentum is followed through, it is found that if the angular 
momentum of the quark motion is $L_q=0$ (corresponding to the lowest energy
quark motion states), then the total angular momentum projection just
equals the angular momentum projection of the junction wave functions,
which in this case is $\pm 1$. Hence the total angular momentum projection
is $\pm 1$ so that $L$ cannot be zero, and should most
likely be 1.

{\bf Exchange symmetry:} 

Exchange symmetry transformations $S_{ij}$ exchange
the positions of the quarks $\bl_i \leftrightarrow \bl_j$.
 Since the physics does not 
depend on the quark position labelling convention, the junction hamiltonian
should be exchange symmetric, as can be seen explicitly in Eq. \ref{ham3}, 
noting that the junction position $\br$ is
not determined by the positions of the quarks.

We now argue that the junction wave functions of (hybrid) baryons 
should transform either totally symmetrically or totally anti--symmetrically
under exchange symmetry. Since the hamiltonian is invariant under
exchange symmetry we have the commutation relation 
$[H_{\mbox{\small flux}},S_{ij}]=0$. Combining this with the Schr\"{o}dinger equation

\beqn\plabel{ex} H_{\mbox{\small flux}}\Psi = V(l_1,l_2,l_3)\Psi \hspace{1cm} 
\mbox{gives} \hspace{1cm} H_{\mbox{\small flux}} (S_{ij}\Psi) =
V(l_1,l_2,l_3) (S_{ij}\Psi)\eeqn
so that $S_{ij}\Psi$ is degenerate in energy with $\Psi$.
Now since the baryon and each of the hybrid baryons $H_i$ have
different energies (except when $l_1=l_2=l_3$) it follows
that  $S_{ij}\Psi$ must be a multiple of $\Psi$, i.e. that
$S_{ij}\Psi = \varsigma \Psi$, where $\varsigma$ is complex number.
Now note that the product of two exchange symmetry transformations is
the identity, i.e. that 

\beqn S_{ij}S_{ij} = 1  \hspace{1cm} \mbox{which implies that}
\hspace{1cm} \varsigma^2 = 1 \eeqn
or $\varsigma=\pm 1$. Hence $S_{ij}\Psi = \pm \Psi$. 

Assume that $S_{12}\Psi = \varsigma\Psi$. We now show\footnote{This result
also follows by noting that $[H_{\mbox{\small flux}},S_{ij}]=0$ implies that
$\Psi$ must be an irreducible representation of the exchange symmetry
group, i.e. totally symmetric, anti--symmetric or mixed symmetry. But since
we already showed that $S_{ij}\Psi = \pm \Psi$, it follows that $\Psi$ is
in either the totally symmetric or anti--symmetric irreducible representation.} that 
$S_{23}\Psi =S_{13}\Psi= \varsigma\Psi$, i.e. that $\Psi$
is either totally symmetric or totally anti--symmetric under label
exchange. This follows by the two identities

\beqna
\lefteqn{S_{12}S_{23}S_{13}S_{23}= 1 \hspace{1cm} \mbox{which implies that}
\hspace{1cm} S_{12}\Psi = S_{13}\Psi  \nonumber } \\ & & \hspace{-0.7cm}
S_{23}S_{12}S_{13}S_{12}= 1 \hspace{1cm} \mbox{which implies that}
\hspace{1cm} S_{23}\Psi = S_{13}\Psi
\eeqna

For each of the hybrid baryons $H_i$, there are hence two varieties: 
the junction wave function is totally symmetric (S) or totally anti--symmetric (A) under quark label exchange, denoted by
$H_i^S$ and $H_i^A$.

{\bf Parity:}

The inversion of all coordinates $\bl_i\rightarrow -\bl_i$ and
$\br\rightarrow -\br$, called ``parity'', is a symmetry of the junction
hamiltonian in Eq. \ref{ham3}. 

$\etahat_{-}$ is a vector in the QQQ plane and is a linear combination
of the $\hat{\bf l}_i$, which span the plane, 
with coefficients which are functions of 
$l_i$. The lengths
$l_i$ remain invariant under parity.
However, $\hat{\bf l}_i \rightarrow - \hat{\bf l}_i$ under parity.
It follows that
$\etahat_{-}$ is {\it odd under parity}. 

The junction wave function
in Eq. \ref{psih} is thus even under parity, since $\etahat_{-}\rightarrow -\etahat_{-}$ and $\br\rightarrow -\br$.
For a low--lying hybrid the quark motion wave function is even under
parity, so that the full hybrid baryon wave function has even parity.

Since quarks are fermions, the wave function 
should be totally antisymmetric under quark label exchange, called
the Pauli principle. 
Since our philosophy is that (hybrid) baryon dynamics is dominated by (non--perturbative) long distance physics, we consider the colour structure of the (hybrid) baryon to be motivated from the long distance limit, i.e. from the strong coupling limit of the hamiltonian formulation of lattice QCD \cite{paton85}. Here, the quarks are sources of triplet colour, which flows along the string connected to the quarks into the junction, where an $\epsilon_{ijk}$ neutralizes the colour. The colour wave function $\epsilon_{ijk}$ is hence totally antisymmetric under exchange of quarks for {\it both} the conventional and hybrid baryon.

This imposes constraints on the combination of 
flavour and non--relativistic spin $S$ of the three quarks that is allowed.
For a totally symmetric hybrid baryon 
junction wave function, the flavour--spin wave
functions must be totally symmetric. This is because we are interested
in the low--lying hybrid baryons which have the 
quark motion wave function in ground state, i.e. totally symmetric. If the
flavour is $\Delta$, which is totally symmetric, this implies that the spin 
must be totally symmetric, i.e. $S=\frac{3}{2}$. Similarly for flavour
$N$ the spin must be $\frac{1}{2}$. For a totally antisymmetric junction
wave function, the flavour--spin wave function must be totally antisymmetric.
For $\Delta$ flavour this implies that the spin must be totally antisymmetric,
which is not realizable. Hence there are no $\Delta$ hybrid baryons with
totally antisymmetric junction wave functions. The $N$ flavour is found to have
spin $\frac{1}{2}$.

The quantum numbers of the lowest--lying states that can be constructed
on the $H_1$ adiabatic surface are indicated in Table \ref{tabqu}.

The total angular momentum ${\bf J} = {\bf L} + {\bf S}$.  Since $L=1$ for   
ground state $H_1$ hybrid baryon, $J=\frac{1}{2},\frac{3}{2}$ for $S=\frac{1}{2}$,
and $J=\frac{1}{2},\frac{3}{2},\frac{5}{2}$ for $S=\frac{3}{2}$.
These assignments are indicated in Table \ref{tabqu}.

One notes from Table \ref{tabqu} that amongst the $H_1^S$ hybrid baryons,
there are $N \frac{1}{2}^+$ and $\Delta \frac{3}{2}^+$
states which have identical quantum numbers to the conventional $N$ and
$\Delta$ baryons.

\begin{table}[t]
\begin{center}
\caption{\small Quantum numbers of low--lying hybrid baryons for the  adiabatic surface $H_1$. In the absense of spin dependent forces all these states are degenerate. $N,\Delta$ are the flavour structure of the wave function (i.e. those of the conventional baryons $N,\Delta$ respectively) and $P$ the parity. } 
\label{tabqu}
\begin{tabular}{|c||l|c|l|}
\hline 
Hybrid Baryon          &  $L$ & $S$ & $(N,\Delta)^{2S+1}J^P$ \\
\hline 
$H_1^S   $ &  1 & $\frac{1}{2},\frac{3}{2}$  &  $N^2 {\frac{1}{2}}^+, \; N^2 {\frac{3}{2}}^+, \; \Delta^4 {\frac{1}{2}}^+, \; \Delta^4 {\frac{3}{2}}^+, \; \Delta^4 {\frac{5}{2}}^+$\\
$H_1^A   $ &  1 & $\frac{1}{2}$  &  $N^2 {\frac{1}{2}}^+, \; N^2 {\frac{3}{2}}^+$\\
\hline 
\end{tabular}
\end{center}
\end{table}

It is interesting to compare our hybrid baryons to the predictions
of the bag model. Out of all the states listed under
$H_1^S$ and $H_1^A$ in Table \ref{tabqu},
only one pair of
 $N^2 {\frac{1}{2}}^+, \; N^2 {\frac{3}{2}}^+$ states have the same
flavour, spin $S$, total angular momentum and parity as 
the low--lying hybrid baryons in the bag model \cite{bag}. 
In fact, for the $H_1^S$
hybrid baryons, the bag model swaps the $N$ and $\Delta$ flavours
from our assignments, keeping other quantum numbers the same.
Both our model and the bag model has seven low--lying hybrid baryons 
\cite{bag}.

\section{Numerical estimate of the hybrid baryon mass}

The difference between the hybrid and conventional baryon adiabatic potentials
(or junction energies) 
as a function of quark positions, $V_{H_1}(l_1,l_2,l_3)-V_B(l_1,l_2,l_3)$,
was determined numerically 
from the first part of Eq. \ref{ex} by using the 
hamiltonian in Eq. \ref{ham3}, and were displayed in ref. \cite{conf}.

Now define the  
hybrid baryon potential as

\beqn\plabel{pot2}
\mbox{\={V}}_{H_1} (l_1,l_2,l_3) \equiv
\mbox{\={V}}_B (l_1,l_2,l_3) + V_{H_1} (l_1,l_2,l_3) - V_B (l_1,l_2,l_3)
\eeqn
where  $\mbox{\={V}}_B (l_1,l_2,l_3)$ is the phenomenologically
successful relativized baryon hamiltonian with Coulomb and 
linear potential terms of ref. \cite{capstick86} (with spin--spin, spin--orbit
and tensor interactions neglected); and the parameters are
also those of ref. \cite{capstick86}.
Note that the Coulomb interaction of the conventional and
hybrid baryon is assumed to be identical.

We solve the Schr\"{o}dinger equation for the hamiltonian in Eq. 
\ref{pot2} with 95 spin--space basis states incorporating $L_q =0,1,2$
harmonic oscillator wave functions for the $J=\frac{1}{2}$ case,
i.e. construct 95 $\times$ 95 dimensional matrices. 
These matrices are subsequently diagonalized.
The differences between the energies for the hybrid and the conventional baryon is then
added to the experimental mass of the lowest baryon, 
taken as the spin--averaged mass of the $N$ and $\Delta$, i.e. 1085 MeV \protect\cite{pdg98}. The first three quark orbital
excitations $L_q=0,1,2$ of hybrid baryons composed of up and down quarks
are found to have masses 1976, 2341 and 2619 MeV respectively.

Hence, for the lowest hybrid baryon level, with the quantum numbers in 
Table \ref{tabqu},
 we obtain that $M_{H_1} - M_B = 891$ MeV, giving a mass estimate of
 $M_{H_1} = 1976$ MeV. 

This mass estimate is substantially higher than other mass estimates in the literature:
$\sim 1.5$ GeV in the bag model \cite{bag} and $1.5\pm 10\%$ GeV in QCD sum rules \cite{zpli}.

There are two crucial assumptions that were made in the early work on (hybrid) meson masses in 
the flux--tube model: the adiabatic motion of quarks and the small oscillation approximation for
flux motion \cite{paton85}. It was later shown that when the adiabatic approximation is lifted,
the masses goes up, and when the small oscillation approximation is lifted, the masses goes down
\cite{paton85}. In our study of (hybrid) baryons we have partially lifted the adiabatic approximation
by working in the centre of mass frame. We have fully lifted the small oscillation approximation.
The effects on the masses of (hybrid) baryons when the various approximations are lifted are the same
as those found for (hybrid) mesons.

In our simulation, we obtain the average values $\sqrt{\langle\rho^2\rangle}=\sqrt{\langle\lambda^2\rangle} = 2.12,\; 2.52$ GeV$^{-1}$
for the low--lying baryon and $H_1$ hybrid baryon respectively.
${\langle\rho^2\rangle}={\langle\lambda^2\rangle}$ is expected since the spatial parts of the
wave functions of the low--lying states
are totally symmetric under exchange symmetry. The hybrid baryon is 20\% larger than the
conventional baryon.

\section{Phenomenology}

The sign of the the Coulomb interaction
is expected to be the same for both conventional and hybrid baryons
\cite{bag}. 
This means that 
the hyperfine interaction has the same sign in both situations, so that 
the $\Delta$ hybrid baryons are always heavier than the $N$ hybrids. 
This implies that only four of the original seven low--lying baryons,
the  $N$ hybrids, are truely low--lying.

We expect {\it a priori} the most phenomenologically interesting decay of the low--lying hybrid baryons to be the P--wave decay to $N\rho$ and $N\omega$, simply because the phase space is favourable and $\rho$ and $\omega$ are easily isolated experimentally. The $N\rho$ decay would be especially relevant to the electro-- and photoproduction of hybrid baryons at TJNAF via the vector meson dominated coupling of the photon to the $\rho$. Indeed, a search for excited $N^*$ resonances with mass $< 2.2$ GeV is currently underway in Hall B \cite{kees}. Given the mass estimate for the low--lying hybrid baryons, the detection of hybrid baryons in $N\rho$ or $N\omega$ is feasible at TJNAF. 
There are also planned experiments in $\pi N$ 
scattering by Crystal Ball E913 at the new D--line at Brookhaven 
with the capability of searching for states
in $N\{\eta,\rho,\omega\}$, which would isolate states in the mass region
$\sim 2$ GeV \cite{bris}. 

The decay $\psi\rightarrow p\bar{p}\omega$ has been observed with a branching ratio of $1.30\pm 0.25\; 10^{-3}$ and $\psi\rightarrow p\bar{p}\eta^{'}$ with branching ratio $9\pm 4\; 10^{-4}$ \cite{pdg98}. Since gluonic hadron production is expected to be enhanced above conventional hadron production in the glue--rich decay of the $\psi$, it is possible that a partial wave analysis of the $p\omega$ or $p\eta^{'}$ invariant masses would yield evidence for hybrid baryons. Future work at BEPC and an upgraded $\tau$--charm factory would be critical here.

\section{Conclusions}

The spin and flavour structure of the low--lying 
hybrid baryons have been specified,
and differ from their structure in the bag model.
Exchange symmetry constrains the spin and flavour of the (hybrid) baryon 
wave function. The orbital angular momentum of the low--lying hybrid baryon
is argued to be unity, with the parity even, contrary to conventional 
baryons where $L=1$ would imply the parity to be odd. 
The low--lying hybrid baryon adiabatic potential and
mass has been estimated numerically. The mass estimate is considerably 
higher than bag model and QCD sum rule estimates.

\end{document}